\documentclass[a4paper, oneside, twocolumn, notitlepage, 10pt]{extarticle_ecoc}
\usepackage{ecoc}

\begin{document}
\selectlanguage{english}    


\title{Data-Driven Modeling of Directly-Modulated Lasers}%


\author{
    S. Hernandez F. \textsuperscript{(1)}, C. Peucheret \textsuperscript{(2)}, O. Jovanovic \textsuperscript{(1)},
    F. Da Ros\textsuperscript{(1)}, D. Zibar \textsuperscript{(1)}
}

\maketitle                  


\begin{strip}
 \begin{author_descr}

   \textsuperscript{(1)} Department of Electrical and Photonics Engineering, Technical University of Denmark, 
   \textcolor{blue}{\uline{shefe@dtu.dk}}

   \textsuperscript{(2)} Univ Rennes, CNRS, FOTON - UMR6082, 22305 Lannion, France.

 \end{author_descr}
\end{strip}

\setstretch{1.1}
\renewcommand\footnotemark{}
\renewcommand\footnoterule{}


\begin{strip}
  \begin{ecoc_abstract}
    The end-to-end optimization of links based on directly-modulated lasers may require an analytically differentiable channel. We overcome this problem by developing and comparing differentiable laser models based on machine learning techniques. 
  \end{ecoc_abstract}
\end{strip}


\section{Introduction}
Directly-modulated lasers (DMLs) are at the core of short-reach communication links 
thanks to their efficiency in terms of power and cost \cite{Mahgerefteh2016Techno-EconomicVCSELs, Huang2021BeyondApplications}. Their potential in terms of transmission distance and line rate is however hindered by their characteristics, such as limited modulation bandwidth, frequency chirping and low extinction ratio. 

Equalization is an effective method to compensate the DML-introduced distortion, but previous solutions have relied on experimental data to drive their models \cite{Wang2019AdvancedTransmission, Huang2021NonlinearSystems}. Further throughput improvements could be achieved by jointly optimizing the transmitter and receiver using end-to-end (E2E) learning, a method that has gained traction as an optimization approach for optical communication systems \cite{Karanov2018End-to-EndCommunications, Jovanovic2021End-to-endLinewidth}. 
This approach usually relies on gradient-based optimization algorithms, that require a differentiable channel model \cite{Karanov2020ConceptModel}. However, the large-signal DML dynamics are governed by nonlinear differential equations for which analytical differentiation cannot be performed \cite{Zhu2018DirectlyLasers} making it challenging to have a differentiable channel. Alternative optimization methods based on reinforcement learning \cite{Aoudia2018End-to-EndModel} and gradient-free optimization \cite{Jovanovic2021Gradient-FreeChannels} have been proposed, but they could be often impractical due to their computational overhead \cite{Yankov:22}. 

A locally-accurate DML surrogate channel enables E2E learning and allows simultaneous optimization of several functions within the communication system \cite{Wang2021TheSimulation}. Previous work using Transformer-based modeling of communication channels has proven the potential of such approaches in the inference of complex dynamical systems, yielding performance gains compared to feed-forward networks and Long-Short Term Memory (LSTMs) \cite{Zhang2022Transformer-BasedSystems, Zhu2023Transformer-basedLink}. 


In this paper, we propose the use of machine learning approaches to learn an accurate differentiable data-driven laser model. The proposed Transformer method is compared to three other common function estimators in dynamical system analysis (Volterra series, time-delay neural networks (TDNNs) and LSTMs). The Transformer model is able to outperform its counterparts while maintaining comparable training and testing time. 

\begin{figure*}[t]
   \centering
    \includegraphics[width=0.925\linewidth]{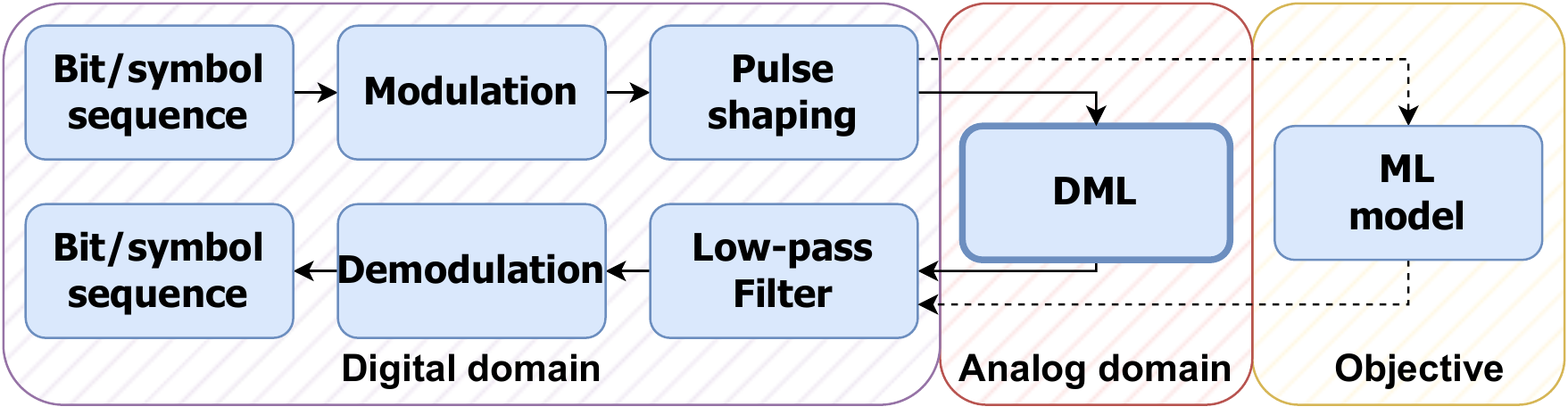}
    \caption{Block diagram of the system under investigation.}
    \label{fig:blockdiag}
\end{figure*}

\section{Data-driven DML modeling}
The overall goal is to emulate the response of any DML laser as closely as possible based only on I/O sequences, as shown in Fig.~\ref{fig:blockdiag}. Transformers are machine learning structures designed for the parallel processing of numerical sequences, avoiding the use of recurrent elements. 
In this work, we propose the use of Convolutional-Attention Transformers (CATs) \cite{Li2019EnhancingForecasting}. CATs make use of convolutions to model the dependencies between temporal sequences. The advantages of this approach are threefold: (i) it limits the amount of past sequence samples used in the prediction, (ii) it is able to capture waveform patterns rather than individual relations between samples; (iii) it takes into account the order of the samples.

The training data acquisition setup is based on numerical simulations obtained from the general laser rate equations \cite{Coldren2012DiodeCircuits} but varying the symbol rate of the driving signal. The solution to the rate equations is obtained using a 5th-order Runge-Kutta (RK4,5) solver. The solution from the solver is then used as ground truth to the CAT, establishing the relation between input modulation current and optical output (power) of the laser. For the data-driven model to be accurate throughout a wide variety of scenarios, the input data must contain a wide range of waveforms and amplitudes, thus providing an exhaustive picture of the behaviour of the laser. This was addressed by switching between two kinds of pulse shapes: super-Gaussian pulses and random pulses, where the latter are sampled from a folded normal distribution $\mathcal{N}(0.5,1)$. The $e^{-2}$ temporal full width $T_{0}$ and the order $n$ of the super-Gaussian pulses are stochastic too, following the folded $\mathcal{N}(0.25T_{sym},T_{sym})$ and uniform $\mathcal{U}(1,6)$ distributions, respectively. The amplitude of the pulses is modulated according to equiprobable 4PAM symbols. The pulses are then min-max normalized and low-pass filtered to avoid out-of-band leakage. The pulse shaping is re-randomized every 8 symbols (with 32 samples per symbol) until completing a 1024-sample sequence of mixed pulse shapes. The training dataset includes $2^{13}$ sequences for a total of $2^{23}$ samples, while the validation set is composed of $2^{17}$ samples. 

\begin{figure}[t]
    \centering
    \includegraphics[width=\linewidth]{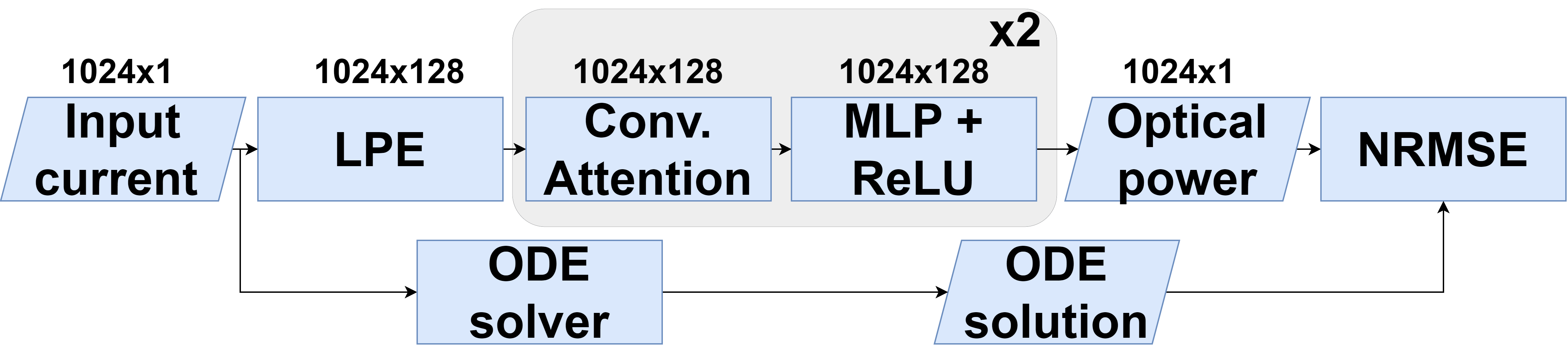}
    \caption{Block diagram of the CAT model setup. }
    \label{fig:convatt}
\end{figure}

The proposed CAT model is based on a decoder-only structure. The network is built around 3 blocks: learned positional embeddings (LPEs), convolutional attention sublayers and 2-layer multi-layer perceptrons (MLP) with ReLU hidden activation, as shown in Fig.~\ref{fig:convatt}. The implemented residual connections are based on the RK2 ordinary differential equation (ODE) Transformer structure \cite{Li2021ODETranslation}, and every sublayer output is then layer-normalized. The reduction of the hidden dimensionality is handled by a linear layer. For the sake of comparison, three additional models have been studied, namely a 2nd order Volterra filter with 16-sample memory, a TDNN and a LSTM \cite{Sutskever2014SequenceNetworks}. The corresponding value for each of the network hyperparameters are gathered in Table~\ref{tab:params}.

\section{Numerical results}

\begin{figure*}[t]
   \centering
    \includegraphics[width=0.95\linewidth]{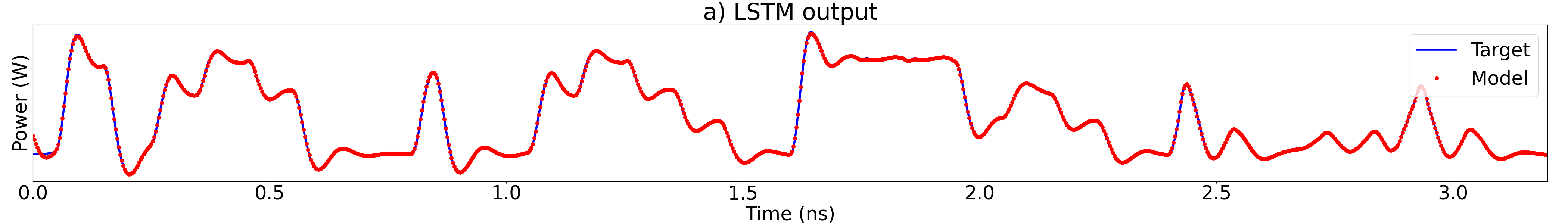}
\end{figure*}
\begin{figure*}[t]
   \centering
    \includegraphics[width=0.95\linewidth]{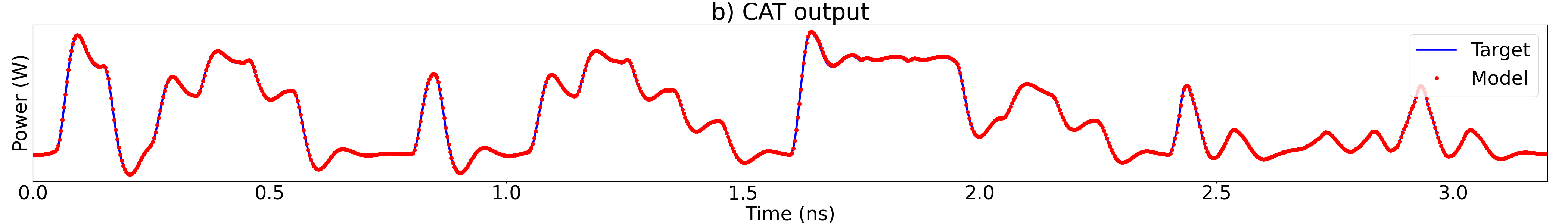}
    \caption{Comparison between the RK4,5 ground truth and test output sequences of a) LSTM and b) CAT}
    \label{fig:seqs}
\end{figure*}

Due to the nature of the laser, the distortion on the optical waveform increases with the symbol rate $R_s$. This effect becomes especially prominent at $R_s$ higher than the relaxation frequency of the laser, $f_R$. It is interesting to focus on these frequencies, where link optimization can have the highest impact. The models were therefore sequentially trained and tested under 5 different symbol rates expressed as fractions of $f_R$ and corresponding to approximately $\{0.1, 0.25, 0.5, 0.75, 1, 1.25\} \cdot f_R$. In every case the training is based on an Adam optimizer with default decay rates $\beta_1 = 0.9$, $\beta_2 = 0.999$ and Normalized Mean Squared Error (NMSE) as a loss function, expressed as NRMSE (taking its square root) for easier interpretation. The laser phase and intensity noise are neglected to avoid setting a lower bound on the MSE performance. All models have been trained for 400 epochs, but only the best test loss is further considered to avoid overfitted results. 

\begin{table}[t]
  \centering
\caption{Model hyperparameters used} \label{tab:params}
\begin{tabular}{|c|c|c|c|}
        \hline  & \textbf{CAT}  & \textbf{TDNN} & \textbf{LSTM}\\
        \hline  \# hidden nodes & 256 & 2048 & 64\\
        \hline \# hidden layers & 2 & 1 & 2 \\ 
        \hline Activ. fun. & ReLU & ReLU & ReLU \\
        \hline  \# MLP sublay. & 2 & 2 & -\\
        \hline  Conv. win. length & 19 & 25 & -\\
        \hline  Embedd. size & 128 & - & -\\
        \hline \# attention heads & 8 & - &- \\ 
        \hline
\end{tabular}
\vspace{-12pt}
\end{table}%

 \begin{figure}[t]
   \centering
    \includegraphics[width=0.95\linewidth]{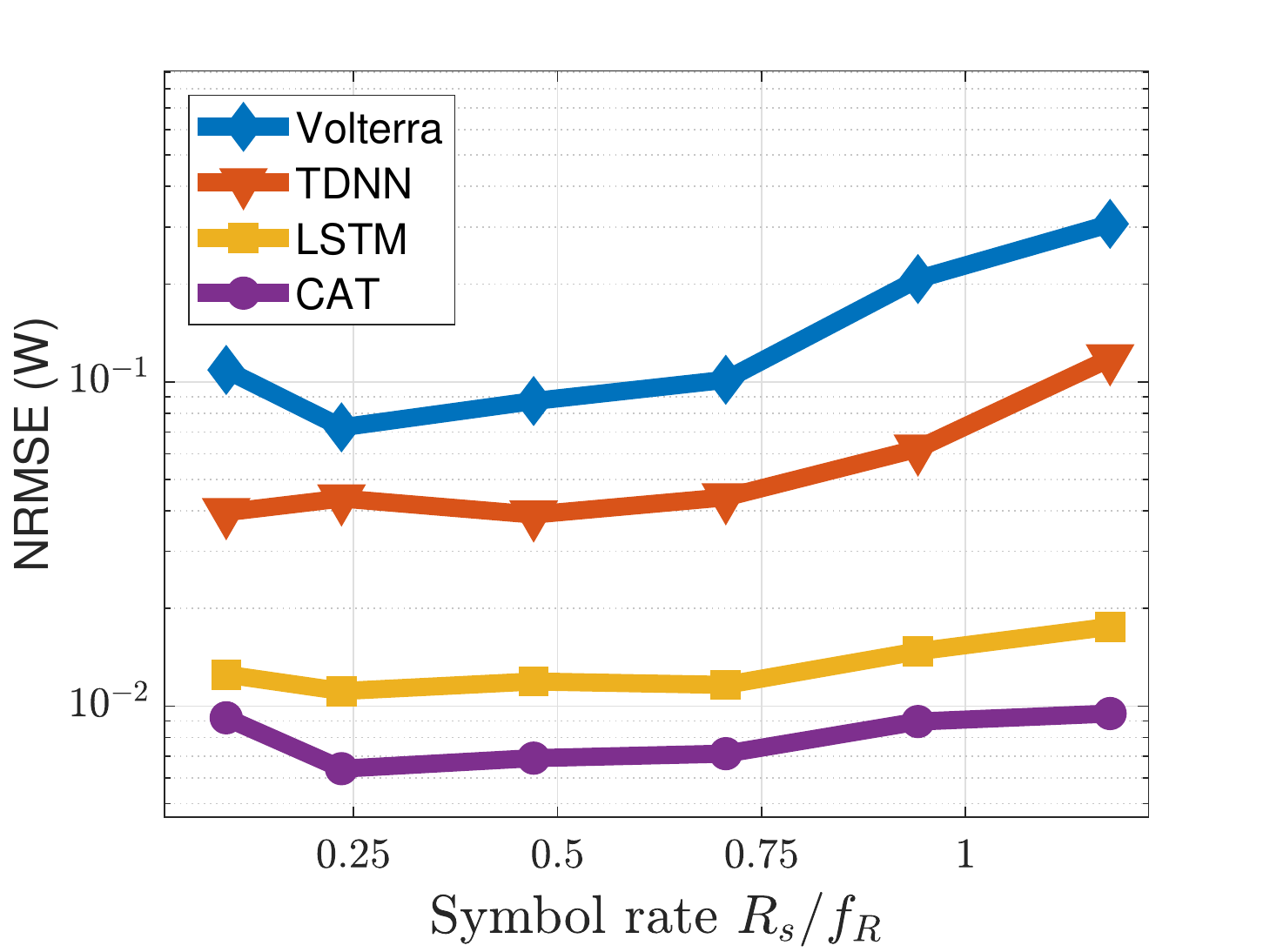}
    \caption{Test NRMSE performance of the proposed models}
    \label{fig:rmse}
\end{figure}

\begin{figure}[t]
    \centering
    \includegraphics[width=0.95\linewidth]{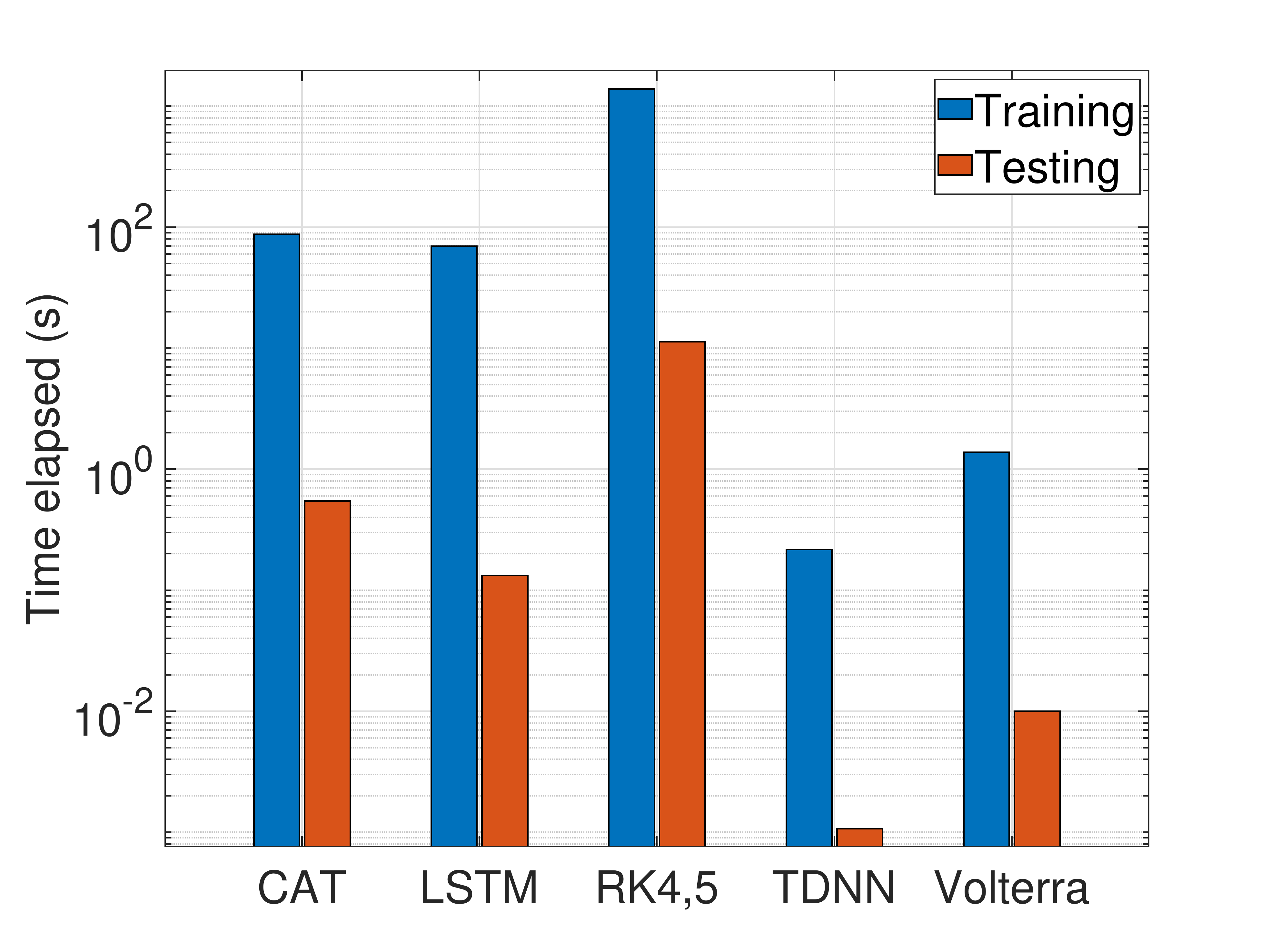}
    \caption{Time elapsed (per epoch) by the presented models}
    \label{fig:times}
\end{figure}

\begin{figure}[t]
    \centering
    \includegraphics[width = 0.95\columnwidth]{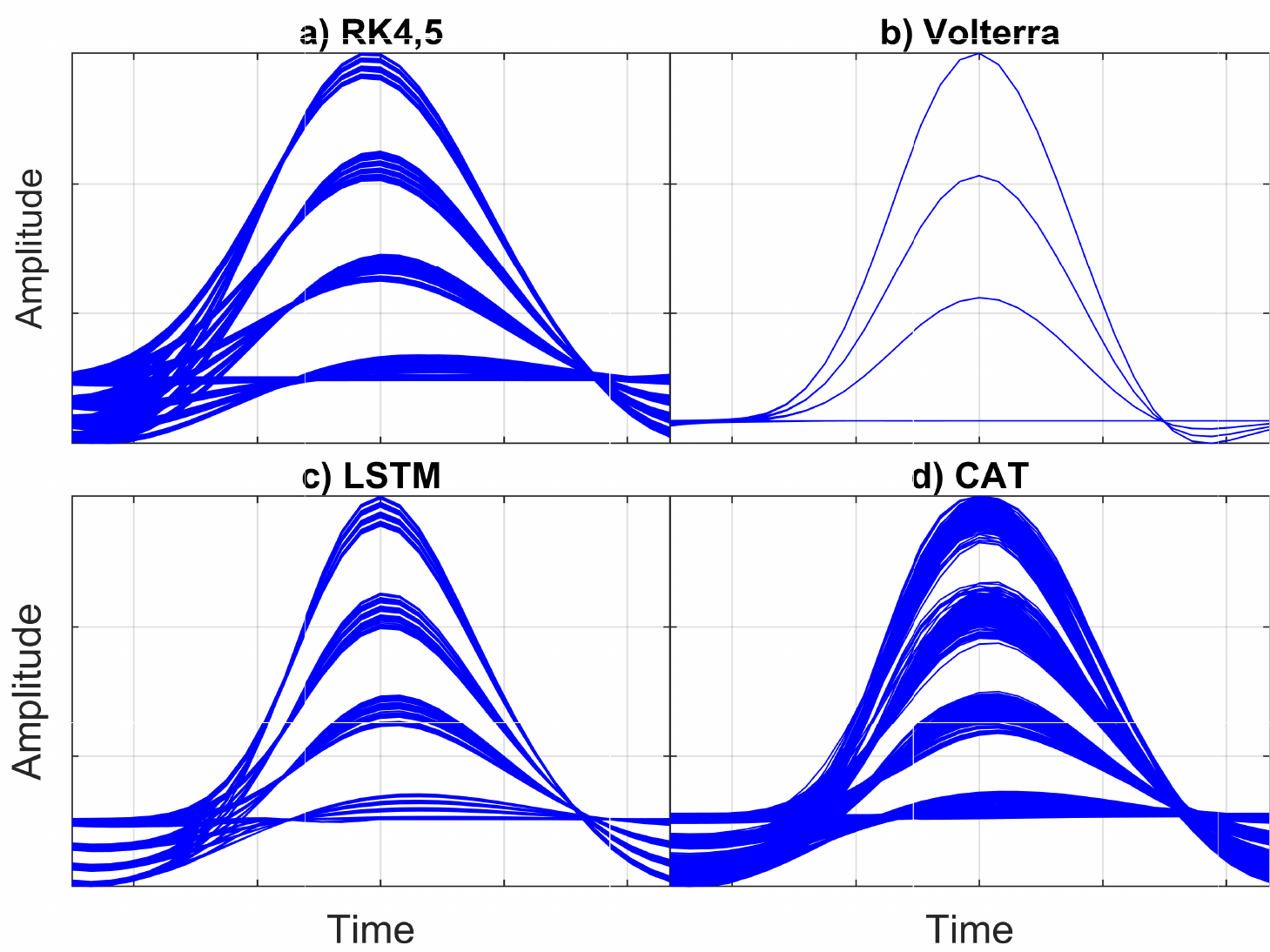}
    \caption{Eye diagram of the proposed models compared to the RK4,5 ODE solver}
    \label{fig:eyes}
\end{figure}
The main attribute of a time-series prediction model is its ability to learn I/O representations. Fig~\ref{fig:seqs} compares the output of the LSTM and CAT models to the RK4,5 solution at $R_s \approx f_R$. Although both figures show high model accuracy, the LSTM struggles to capture the first few samples of the sequence. This is probably due the high reliance of LSTMs in their memory mechanism, that limits its performance when little temporal context is provided.  A similar trend is shown in Fig.~\ref{fig:rmse}, where NRMSE is shown as a function of the symbol rate. Throughout the analyzed bandwidth, the CAT outperforms its counterparts and falls under the $10^{-2}$ mark that sets the 1\% error threshold. The trend of the 4 curves hints the correlation between $R_s$ and the waveform distortion introduced, i.e. as the symbol duration becomes shorter it becomes increasingly difficult to match the input and output sequences for all models. It must be noted that, even though the CAT has more training parameters, its parallelization potential makes its training and inference time per sequence comparable to the LSTM. This can be seen in Fig.~\ref{fig:times}, where the time elapsed to process both the training and testing sequences on a Nvidia A100 GPU is compared. It is also evident how the training and testing time per epoch of all the proposed approaches is at least an order of magnitude faster than the ODE solver generating the training and testing data. Looking at the eye diagrams at $R_s \approx f_R$ for Gaussian input pulses in Fig.~\ref{fig:eyes}, the trend reveals a more contrasted picture than the NRMSE alone. Even if all 4 models show reasonable convergence compared to the ODE case (with the exception of the TDNN, that was omitted due to its poor performance), the Volterra filter and the LSTM show a consistent performance through the $2^{10}$ symbols shown, while the CAT seems more sensitive to small variations of position and amplitude in the samples. This could be due to the positional encoding in the model, that alters the input to the network based on the position of the sample, even if its value remains constant. This drawback may however be less relevant in real scenarios, where noisy input data would affect the resulting output waveform to some degree. In terms of capturing the true width of the output pulse, the CAT shows a slightly better tracking than its counterparts, which tend to shorten the modulated pulse duration.

\section{Conclusions}
A data-driven differentiable surrogate for directly-modulated lasers was proposed. We show that the Transformer model is able to accurately predict the laser response while maintaining similar inference time compared to other time-series approaches. Our results can enable the joint optimization of directly-modulated systems without relying on experimental data or online gradient approximations.

\section{Acknowledgements}
This work was financially supported by the ERC-CoG FRECOM project (no. 771878) and the Villum YIP OPTIC-AI project (no. 29334).

\newpage
\newpage
\printbibliography

\vspace{-4mm}

\end{document}